%% file: Borsboom.Fahdzyana.ea.VPPC20.tex
\documentclass[conference]{IEEEtran}
\renewcommand{\baselinestretch}{0.96}
\usepackage{cite}

\ifCLASSINFOpdf
\usepackage[pdftex]{graphicx}
\graphicspath{Matlab}
\DeclareGraphicsExtensions{.pdf,.jpeg,.png}
\else
\fi
\usepackage{array}

\ifCLASSOPTIONcompsoc
\usepackage[caption=false,font=normalsize,labelfont=sf,textfont=sf]{subfig}
\else
\usepackage[caption=false,font=footnotesize]{subfig}
\fi
\usepackage{url}
\usepackage{amsthm}
\usepackage{amssymb}
\newcounter{thm}

\theoremstyle{definition}

\newtheorem{prob}[thm]{Problem}

\usepackage[pdftex]{graphicx}
\graphicspath{Matlab}
\DeclareGraphicsExtensions{.pdf,.jpeg,.png}
\usepackage{url}
\usepackage{booktabs}
\usepackage{lettrine}

\usepackage{tikz,pgfplots}

\input{def}
\usetikzlibrary{shapes,arrows,positioning}
\usetikzlibrary{calc}
\usetikzlibrary{plotmarks}
\usepackage{fixmath}
\usepackage{amsmath}
\usepackage{amssymb}
\usepackage{mathrsfs}

\usepackage{algorithm}
\usepackage[noend]{algpseudocode}

\makeatletter
\def\BState{\State\hskip-\ALG@thistlm}
\makeatother

\floatname{algorithm}{Algorithm}

\usepackage{units}
\usepackage{mathtools}
\usepackage{dsfont}
\usepackage{booktabs}
\usepackage{bbm}
\usepackage{soul}

\newif\ifmargincomments 
\margincommentstrue

\ifmargincomments

\else

\fi

\begin{document}
	%
	\title{
	Time-optimal Control Strategies for Electric Race Cars with Different Transmission Technologies
	}
	%
	%
	%
	
	\author{Olaf Borsboom, Chyannie A.\ Fahdzyana, Mauro Salazar, Theo Hofman
		\thanks{M. Salazar is with the Control Systems Technology group, Eindhoven University of Technology (TU/e), Eindhoven, MB 5600, The Netherlands, \tt\small m.ru.salazar.villalon@tue.nl}
	}
	
	%
	%

	\markboth{IEEE Vehicular Power and Propulsion, October 2020}%
	{Shell \MakeLowercase{\textit{et al.}}: Bare Demo of IEEEtran.cls for IEEE Journals}
	%



	\maketitle
	
	\begin{abstract}
	This paper presents models and optimization methods to rapidly compute the achievable lap time of a race car equipped with a battery electric powertrain. Specifically, we first derive a quasi-convex model of the electric powertrain, including the battery, the electric machine, and two transmission technologies: a single-speed fixed gear and a continuously variable transmission (CVT). Second, assuming an expert driver, we formulate the time-optimal control problem for a given driving path and solve it using an iterative convex optimization algorithm. Finally, we showcase our framework by comparing the performance achievable with a single-speed transmission and a CVT on the Le Mans track. Our results show that a CVT can balance its lower efficiency and higher weight with a higher-efficiency and more aggressive motor operation, and significantly outperform a fixed single-gear transmission.
	\end{abstract}
	\begin{IEEEkeywords}
		Electric vehicles, convex optimization. 
	\end{IEEEkeywords}

	%
	\IEEEpeerreviewmaketitle
	
	\input{chapters/introduction}

	\input{chapters/methodology}
	\input{chapters/results}
	\input{chapters/conclusion}
	
	\section*{Acknowledgment}
	\noindent
	We thank Dr.\ I.\ New for proofreading this paper and Dr.~S.~Singh for the fruitful discussions.
	
	

	
	
	%
	
	\bibliographystyle{IEEEtran}        

	\bibliography{../../../Bibliography/main,../../../Bibliography/SML_papers}

\end{document}

%% file: def.tex
\newcommand{\dds}{\frac{\textnormal{d}}{\textnormal{d}s}}
\newcommand{\dtds}{\frac{\textnormal{d}t}{\textnormal{d}s}}

%% file: chapters/introduction.tex
\section{Introduction}\label{sec:introduction}
\lettrine{O}{ver} the last few years, the electrification of vehicle powertrain systems has gained significant interest. This trend is not only visible in commercial and passenger vehicles, but also in the racing community, with the introduction of the fully-electric Formula E racing class and the strong hybridization of the Formula~1 power unit taking place in 2014~\cite{FIA}. In contrast to commercial vehicles and heavy-duty trucks where the design and control goals are aimed at reducing the energy consumption, the most important performance indicator for race cars is the lap time.
In this context, the control strategies managing the energy deployment of the vehicle on the racetrack as well as the design of its powertrain have a significant impact on the achievable lap time and must be carefully optimized. This calls for methods to compute the minimum-lap-time control strategies and concurrently perform optimal powertrain design studies for battery electric race cars.

\begin{figure}[!t]
	\centering
	\includegraphics[ width=\columnwidth]{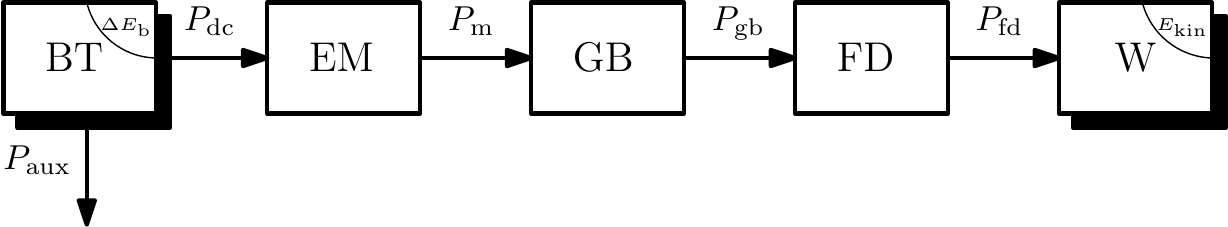}
	\caption{Schematic layout of the considered electric race car powertrain that consists of a battery pack (BT), an electric machine (EM), a transmission~(GB) consisting of a single gear-ratio (SR) or a continuously variable transmission (CVT), and a final drive reduction gear (FD) connected to the wheels (W). The arrows indicate the power flows between the components.}
	\label{fig:PTlayout}
\end{figure}

This paper presents a convex modeling and optimization framework to efficiently compute the minimum-lap-time control strategies for the battery electric race car shown in Fig.~\ref{fig:PTlayout}. This framework allows to efficiently compute the time-optimal control strategies and characterize the impact of powertrain design choices on the achievable lap~time.
\subsubsection*{Related literature}
The problem studied in this paper pertains to two main research streams. The first line consists of offline control strategies for the fuel-optimal energy management of hybrid electric vehicles. In general, such non-causal approaches are based on Pontryagin's minimum principle \cite{SciarrettaGuzzella2007, NgoHofmanEtAl2012, TangRizzoniEtAl2015, RitzmannChristonEtAl2019}, dynamic programming \cite{PerezPilotta2009, ElbertEbbesenEtAl2013, LinKangEtAl2001}, and convex optimization~\cite{ElbertNueeschEtAl2014, MurgovskiJohannessonEtAl2015, RobuschiSalazarEtAl2019}.
The second research line is related to the time-optimal control of racing vehicles such as the hybrid electric Formula~1 car.
The majority of the existing research in this field optimizes the velocity profile and driving trajectory simultaneously~\cite{Casanova2000,LotEvangelou2013,LimebeerPerantoniEtAl2014, LimebeerRao2015}, or separately solve the energy management problem offline~\cite{EbbesenSalazarEtAl2018,SalazarElbertEtAl2017,SalazarDuhrEtAl2019} and control the power unit in real-time~\cite{SalazarBalernaEtAl2017,SalazarBalernaEtAl2018}, using convex optimization.
To the best of the authors' knowledge there are no optimization methods for battery electric race cars explicitly accounting for the technology, sizing and control of the transmission.

\subsubsection*{Statement of contributions}
Against this background, this paper presents a computationally-efficient optimization framework for battery electric race cars equipped with two types of vehicle transmissions, namely, a single gear-ratio (SR) or a continuously variable transmission (CVT), whereby we optimize the design of the SR and the control of the CVT.
First, we derive an almost-convex model of the powertrain shown in Fig.~\ref{fig:PTlayout}.
Second, we formulate the time-optimal control problem and solve it with an iterative procedure based on second-order conic programming.
Finally, we showcase our algorithmic framework with a case study on the Le Mans track.

\subsubsection*{Organization}
The remainder of this paper is structured as follows: Section~\ref{sec:methodology} presents the time-optimal control problem and an effective solution algorithm based on convex optimization. Section~\ref{sec:results} presents numerical results for both an SR and a CVT-equipped powertrain on the LeMans track. Finally, we draw the conclusions in Section~\ref{sec:conclusion}.

%% file: chapters/methodology.tex
\section{Methodology}\label{sec:methodology}
In this section, we identify an almost-convex model of the considered vehicle and powertrain, formulate the minimum-lap-time control problem and devise an iterative approach to solve it.
The first two steps are inspired by the methods applied to Formula~1 in~\cite{EbbesenSalazarEtAl2018}, but differ in terms of powertrain topology and the explicit characterization of the transmission technology and its impact on the powertrain performance.

Fig.~\ref{fig:PTlayout} shows a schematic layout of the powertrain of the vehicle under consideration. It is propelled by an electric motor (EM), which converts electric energy from the battery pack to kinetic energy. Hereby, we consider two types of transmission: an SR transmission and a CVT. A final reduction transfers the resulting mechanical energy to the wheels.
The input variables are the motor power ${P_{\mathrm{m}}}$ and, in the case of the CVT-equipped powertrain, also the transmission ratio ${\gamma}$. We treat the transmission ratio as a design variable when considering the SR.
The state variables are the kinetic energy of the vehicle ${E_{\mathrm{kin}}}$ and the amount of battery energy used since the start of the lap ${\mathrm{\Delta} E_{\mathrm{b}}}$.
The powertrain controller cannot directly influence the driving path, which is rather set by the expert driver's steering input, hence we condense the  3D characteristics of the racetrack and the driven path in a 1D maximum-velocity profile $v_\mathrm{max}(s)$. This profile can either be measured or pre-computed, and depends on the longitudinal position of the car on the racetrack $s$.

\subsection{Minimum-lap-time Objective}
We define the time-optimal control problem for one free-flow race lap in space-domain. This way, we can directly implement position-dependent parameters such as the maximum speed profile, and the problem has a finite-horizon.
The objective of the control problem is minimizing lap time $T$, i.e.,
\small
\begin{equation} \label{eq:objective}
	\text{min } T = \text{min} \int_{0}^{S} {\dtds(s)} \mathrm{d}s,
\end{equation}
\normalsize
where $S$ is the length of the track and $\dtds(s)$ is the lethargy. This can be expressed as the \textit{time per driven distance}, which is the inverse of the vehicle's speed $v(s)$:
\small
\begin{equation}\label{eq:lethargy}
	\dtds(s) = \frac{1}{{v(s)}}.
\end{equation}
\normalsize
However, since $\dtds(s)$ and $v(s)$ are both optimization variables, (\ref{eq:lethargy}) is a non-convex constraint. Rearranging and relaxing (\ref{eq:lethargy}) results in
\small
\begin{equation*}
	\dtds(s) \cdot {v(s)} \geq 1,
\end{equation*} 
\normalsize
which is a geometric mean expression and can be written as the second-order conic constraint
\small
\begin{equation}\label{eq:lethargycone}
	\dtds(s) + v(s) \geq 
	\left \| \begin{matrix}
	2 \\
	\dtds(s) - v(s)
	\end{matrix}   \right \|_2.
\end{equation}
\normalsize
Since from the objective \eqref{eq:objective}, it is optimal to minimize the lethargy $\dtds(s)$, the solver will converge to a solution where constraint \eqref{eq:lethargycone} holds with equality~\cite{EbbesenSalazarEtAl2018}.

In the following sections, the longitudinal and powertrain dynamics and constraints will be expressed in space-domain and relaxed to a convex form whenever necessary.
Since the space-derivative of energy is force, we will ultimately define the model of the powertrain in terms of forces. Thereby, power and force are related as
\small
\begin{equation*}
F = \frac{P}{v} = P \cdot \dtds,
\end{equation*}
\normalsize
which can be used in post-processing to compute the optimal power $P^\star$ from the optimal force $F^\star$ and speed $v^\star$.

\subsection{Longitudinal Vehicle Dynamics}
This section derives the longitudinal vehicle dynamics and expresses them in a convex form in space domain.
In order to connect the vehicle dynamics and the objective, we define the physical constraint on the kinetic energy
\small
\begin{equation*}
	E_{\mathrm{kin}} (s) = {m_{\mathrm{tot}}}   \cdot v(s)^2 / {2}, 
\end{equation*}
\normalsize
where $m_{\mathrm{tot}}$ is the total mass of the vehicle. To ensure convexity, we have to relax the kinetic energy to
\small
\begin{equation}\label{eq:kineticenergy}
	E_{\mathrm{kin}} (s) \geq {m_{\mathrm{tot}}} \cdot v(s)^2 / {2}.
\end{equation}
\normalsize
Considering the scope of this study, we disregard lower level longitudinal, lateral, and vertical vehicle dynamics and decide to model the vehicle as a point mass. Applying Newton's Second Law in space-domain to this point mass gives
\small
\begin{equation}\label{eq:newtons2ndlaw}
\dds E_{\mathrm{kin}} (s) = {F_{\mathrm{p}}(s)} - {F_{\mathrm{d}}(s)},
\end{equation}
\normalsize
where ${F_{\mathrm{p}}(s)}$ is the propulsion force and ${F_{\mathrm{d}}(s)}$ is the drag force. The drag force is the sum of the aerodynamic drag force, the gravitational force,  and the rolling resistance force as
\small
\begin{multline}\label{eq:dragforce}
	F_{\mathrm{d}}(s) = {c_\mathrm{d} \cdot A_\mathrm{f}\cdot \rho}\cdot E_\mathrm{kin}(s) / m_{\mathrm{tot}} + \\ m_{\mathrm{tot}} \cdot g \cdot (\sin(\theta(s)) + c_{\mathrm{r}} \cdot \cos(\theta(s))),
\end{multline}
\normalsize
where $c_{\mathrm{d}}$ is the drag coefficient, $A_{\mathrm{f}}$ is the frontal area of the vehicle, $\rho$ is the density of air, $g$ is earth's gravitational constant, $\theta(s)$ is the inclination of the track, and $c_{\mathrm{r}}$ is the rolling friction coefficient.
The propulsive force is equal to
\small
\begin{equation*}
	F_{\mathrm{p}}(s) = 
		\begin{cases}
			\eta_{\mathrm{fd}} \cdot F_{\mathrm{gb}}(s) - F_{\mathrm{brk}}(s) & \mbox{if }   F_{\mathrm{gb}}(s) \geq 0 \\
			\frac{1}{\eta_{\mathrm{fd}}} \cdot  F_{\mathrm{gb}}(s) - F_{\mathrm{brk}}(s)  & \mbox{if }   F_{\mathrm{gb}}(s) < 0,
		\end{cases}
\end{equation*}
\normalsize
where $F_{\mathrm{gb}}(s)$ is the force on the secondary axle of the transmission and $F_{\mathrm{brk}}(s)$ is the force excited by the mechanical brakes. Since $F_{\mathrm{brk}}(s) \geq 0$, the propulsion force can be relaxed to
\small
\begin{equation}\label{eq:propulsiveforce}
		F_{\mathrm{p}}(s) \leq \min \left(\eta_{\mathrm{fd}} \cdot F_{\mathrm{gb}}(s),   F_{\mathrm{gb}}(s) / \eta_{\mathrm{fd}} \right).
\end{equation}
\normalsize
We condense the grip limitations as well as the lateral vehicular dynamics during cornering into the maximum kinetic energy constraint
\small
\begin{equation}\label{eq:maxkineticenergy}
	{E_{\mathrm{kin}}(s)} \leq E_{\mathrm{kin,max}}(s) = {m_{\mathrm{tot}}}\cdot v_\mathrm{max}^2(s)/{2}.
\end{equation}
\normalsize
As mentioned at the beginning of this section, the maximum speed profile can be either pre-computed or measured, capturing the way the expert driver is \emph{feeling} the car.
Finally, considering a free-flow racing lap, we enforce periodicity on the speed of the car with
\small
\begin{equation}\label{eq:Ekinperiod}
	E_{\mathrm{kin}}(0) = E_{\mathrm{kin}}(S).
\end{equation}
\normalsize

\subsection{Transmission}
The speed of the motor is given by
\small
\begin{equation}\label{eq:ratioconstraint}
	\omega_{\mathrm{m}}(s) = \gamma(s) \cdot v(s) \cdot \gamma_{\mathrm{fd}} / r_{\mathrm{w}} ,
\end{equation}
\normalsize
where $\gamma(s)$ is the ratio of the transmission, $\gamma_{\mathrm{fd}}$ is the fixed transmission ratio of the final drive,  and $r_{\mathrm{w}}$ is the radius of the wheels. For the transmission ratio, we define
\small
\begin{equation}\label{eq:ratiolimit}
\gamma(s) 
\begin{cases}
= \gamma_1 & \text{if SR}   \\
\in [\gamma_{\mathrm{min}}, \gamma_{\mathrm{max}}] & \text{if CVT},  
\end{cases}
\end{equation}
\normalsize
where $\gamma_1 > 0$ is the fixed ratio of the SR, and $\gamma_{\mathrm{min}} > 0 $ and $\gamma_{\mathrm{max}} > 0$ are the lower and upper limit of the CVT ratio, respectively. Assuming a constant transmission efficiency $\eta_{\mathrm{gb}}$, we model and relax the transmission force similar to (\ref{eq:propulsiveforce}):
\small
\begin{equation}\label{transmissionforcebalance}
	F_{\mathrm{gb}}(s) \leq \min \left(\eta_{\mathrm{gb}} \cdot F_{\mathrm{m}}(s),   F_{\mathrm{m}}(s) / \eta_{\mathrm{gb}} \right).
\end{equation}
\normalsize

\subsection{Electric Motor} 
In this section, we derive two models of the EM: a speed-independent convex model and a more precise speed-dependent model. The first, coarser model will be used to determine an initial guess for the optimal speed profile, while the latter, speed-dependent model will be leveraged to capture the EM efficiency more accurately. Both models will be instrumental to the iterative solution algorithm based on convex optimization presented in detail in Section~\ref{sec:optprob}.

\begin{figure}[!t]
	\centering
	\includegraphics[width=\columnwidth]{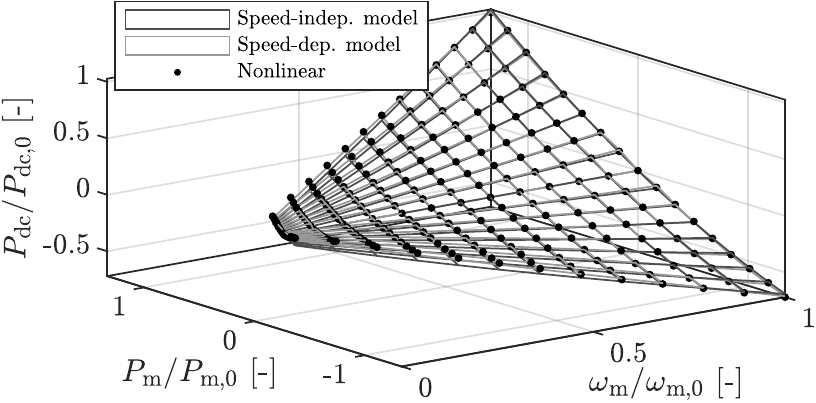}
	\caption{A speed-independent and a speed-dependent electric motor model. The RMSE of the speed-independent model is 2.3\%. The RMSE of the speed-dependent model is 0.23\%.}
	\label{fig:EMmodel}
\end{figure}

First, we approximate the losses with a quadratic function. This way, the electrical EM power is equal to
\small
\begin{equation*}
{P_{\mathrm{dc}}}(s) = \alpha_{\mathrm{m}} \cdot {P_{\mathrm{m}}(s)^2} + {P_{\mathrm{m}}(s)},
\end{equation*}
\normalsize
where $\alpha_{\mathrm{m}} \geq 0 $ is an efficiency parameter subject to identification. Converting this equation to forces leads to
\small
\begin{equation*}
F_{\mathrm{dc}}(s)/v(s) \geq {\alpha_{\mathrm{m}} \cdot F_{\mathrm{m}}(s)^2} + {F_{\mathrm{m}}(s)/v(s)}.
\end{equation*}
\normalsize
We can rewrite this as the geometric mean expression
\small
\begin{equation*}
\dtds(s)\cdot (F_{\mathrm{dc}}(s) - F_{\mathrm{m}}(s)) \geq {\alpha_{\mathrm{m}} \cdot F_{\mathrm{m}}(s)^2},
\end{equation*}
\normalsize
and subsequently as a second order conic constraint as
\small
\begin{equation}\label{eq:EMcone}
{\dtds(s)} + F_{\mathrm{dc}}(s) - F_{\mathrm{m}}(s) \geq 
\left \| \begin{matrix}
2 \cdot \sqrt{\alpha_{\mathrm{m}}} \cdot F_{\mathrm{m}}(s) \\
{\dtds(s)} - F_{\mathrm{dc}}(s) + F_{\mathrm{m}}(s)
\end{matrix}   \right \|_2,
\end{equation}
\normalsize
which will hold with equality in the case where the solver converges to a time-optimal solution and the battery energy is limited~\cite{EbbesenSalazarEtAl2018}.
Fig.~\ref{fig:EMmodel} shows the regression of the nonlinear model in which $\alpha_{\mathrm{m}}$ is identified. For reasons of confidentiality, all the data have been normalized.
In order to capture the impact of speed on the EM efficiency, the losses of the EM can be fitted as a function of mechanical power and speed. We can approximate the nonlinear model in a convex manner using a positive semi-definite approach. In that case, the losses of the EM $P_{\mathrm{m,loss}}(s)$ are equal to
\small
\begin{equation*}
P_\mathrm{m,loss}(s) = x^\top Q x,
\end{equation*}
\normalsize
where $x = \begin{bmatrix}
1  & {\omega_{\mathrm{m}}(s)} & {P_{\mathrm{m}}(s)}
\end{bmatrix}^\top$, ${P_{\mathrm{m}}}(s)$ is the mechanical motor power, and $Q$ is a symmetric and positive semi-definite matrix. The values of $Q$ are determined using semi-definite programming solvers. The electrical power is
\small
\begin{equation*}
{P_{\mathrm{dc}}}(s) = {P_{\mathrm{m}}}(s) + {P_{\mathrm{m,loss}}}(s).
\end{equation*}
\normalsize
Fig.~\ref{fig:EMmodel} shows the resulting speed-dependent electric motor model.
We relax the constraint to derive an expression for $F_{\mathrm{dc}}(s)$:
\small
\begin{equation}\label{eq:motorconstraint}
	F_{\mathrm{dc}}(s) \geq F_{\mathrm{m}}(s) +  \left({{x(s)}^\top Q {x(s)}}\right)/{v(s)},
\end{equation}
\normalsize
where ${x} = \begin{bmatrix}
1  & {\omega_{\mathrm{m}}}(s) & {F_{\mathrm{m}}(s)\cdot v(s)}
\end{bmatrix}^\top$ and $F_{\mathrm{m}}(s)$ is the force translation of mechanical motor power.
The relaxations imposed on the constraints will hold with equality when an optimal solution is attained and the battery energy is limited. However, the constraint above is not completely convex, since $\omega_{\mathrm{m}}(s)$, $F_{\mathrm{m}}(s)$ and $v(s)$ are all optimization variables.
Yet, convexity can be achieved if the vehicle velocity is given, an issue that will be resolved in Section~\ref{sec:optprob}.

The upper and lower bounds of the EM power can be divided into a maximum-torque and maximum-power region. The maximum torque limit is approximated by a linear constraint
\small
\begin{equation*}
	P_{\mathrm{m}}(s) \in [-T_{\mathrm{max}} \cdot \omega_{\mathrm{m}}(s), T_{\mathrm{max}} \cdot \omega_{\mathrm{m}}(s)],
\end{equation*} 
\normalsize
where $T_{\mathrm{max}}$ is a constant maximum torque value. The maximum power region is captured by the affine function
\small
\begin{equation*}
	P_{\mathrm{m}}(s) \in [- c_{\mathrm{m,1}} \cdot \omega_{\mathrm{m}}(s) - c_{\mathrm{m,2}}, c_{\mathrm{m,1}} \cdot \omega_{\mathrm{m}}(s) + c_{\mathrm{m,2}}],
\end{equation*}
\normalsize
where $c_{\mathrm{m,1}} \leq 0 $ and $c_{\mathrm{m,2}} \geq 0$ are parameters subject to identification. The translation to forces results in
\small
\begin{equation}\label{eq:torquelimit}
	F_{\mathrm{m}}(s) \in \left[- \frac{\gamma(s) \cdot T_{\mathrm{max}}}{r_{\mathrm{w}}}, \frac{\gamma(s) \cdot T_{\mathrm{max}}}{r_{\mathrm{w}}} \right],
\end{equation}
\normalsize
and
\small
\begin{equation}\label{eq:powerlimit}
	F_{\mathrm{m}}(s) \in \left[-1,1\right]\cdot\left(\frac{c_{\mathrm{m,1}} \cdot \gamma(s)}{r_{\mathrm{w}}} + c_{\mathrm{m,2}} \cdot \dtds(s) \right),
\end{equation}
\normalsize
where $r_{\mathrm{w}}$ is the radius of the wheels.
Finally, the rotational speed of the motor
cannot exceed the maximum speed $ \omega_{\mathrm{m,max}}$,
which can be expressed as
\small
\begin{equation}\label{eq:speedlimit}
\begin{aligned}
&\gamma(s) \leq \omega_{\mathrm{m,max}}\cdot r_{\mathrm{w}}\cdot \dtds(s).\\
\end{aligned}
\end{equation}
\normalsize

\subsection{Battery Pack}
In this section, we derive a model of the battery dynamics. 
The electric power at the terminals of the battery pack is
\small
\begin{equation*}
	P_{\mathrm{b}}(s) = {P_{\mathrm{dc}}(s)} + {P_{\mathrm{aux}}},
\end{equation*}
\normalsize
where $P_{\mathrm{aux}}$ is a constant auxiliary power. Converting this constraint to forces gives
\small
\begin{equation}
	F_{\mathrm{b}}(s) = { F_{\mathrm{dc}}(s)} + {P_{\mathrm{aux}}}\cdot \dtds(s).
\end{equation}
\normalsize
The internal battery power $P_\mathrm{i}(s)$, which causes the actual change in the battery state of energy $E_{\mathrm{b}}(s)$, is approximated by
\small
\begin{equation*}
		P_{\mathrm{i}}(s) = {\alpha_{\mathrm{b}} \cdot P_{\mathrm{b}}(s)^2} + {P_{\mathrm{b}}(s)},
\end{equation*}
\normalsize
where, similar to~\cite{EbbesenSalazarEtAl2018}, the efficiency parameter $\alpha_{\mathrm{b}}$ is determined with a quadratic regression of the discharge measurement data with an RMSE of 2.3\%.
Similar as for the EM, relaxing this constraint and converting it to forces results in the second order conic constraint
\small
\begin{equation}\label{eq:batterycone}
{\dtds(s)} + F_{\mathrm{i}}(s) - F_{\mathrm{b}}(s) \geq 
\left \| \begin{matrix}
2 \cdot \sqrt{\alpha_{\mathrm{b}}} \cdot F_{\mathrm{b}}(s) \\
{\dtds(s)} - F_{\mathrm{i}}(s) + F_{\mathrm{b}}(s)
\end{matrix}   \right \|_2,
\end{equation}
\normalsize
which, again, will hold with equality when an optimal solution is retrieved and the battery energy is limited~\cite{EbbesenSalazarEtAl2018}.

Finally, the battery energy consumption from the beginning of the lap $\mathrm{\Delta} E_{\mathrm{b}}(s) = E_{\mathrm{b}}(0) - E_{\mathrm{b}}(s)$ is modeled as
\small
\begin{equation}\label{eq:Ebstate}
\dds\mathrm{\Delta} E_\mathrm{b}(s) = F_{\mathrm{i}}(s).
\end{equation}
\normalsize
The available battery energy $E_{\mathrm{b,0}}$  is equally divided by the number of laps $N_\mathrm{laps}$ as $\mathrm{\Delta} E_{\mathrm{b,max}}= {E_{\mathrm{b,0}}} / {N_{\mathrm{laps}}}$ so that
\small
\begin{equation}\label{eq:Ebmin}
	\mathrm{\Delta} E_{\mathrm{b}}(S) \leq \mathrm{\Delta} E_{\mathrm{b,max}}.
\end{equation}
\normalsize
At the beginning of the lap it holds
\small
\begin{equation} \label{eq:Eb0}
\mathrm{\Delta} E_{\mathrm{b}}(0) = 0.
\end{equation}
\normalsize

\subsection{Minimum-lap-time Optimization Problem}\label{sec:optprob}
We present the time-optimal control problem for state and control variables $x = \{E_{\mathrm{kin}}, \mathrm{\Delta} E_{\mathrm{b}} \} $ and $u = \{F_{\mathrm{m}}, \gamma \}$, respectively, and propose a solution method based on a convex speed-independent approximation and a convex formulation with partial application of an estimate of the velocity profile.
\begin{prob}[Nonlinear Problem]\label{prob:main}
	\emph{The minimum-lap-time control strategies are the solution of}
	\small
	\begin{equation*}
		\underset{x,u}{\text{min}}  \int_{0}^{S} {\dtds(s)}\mathrm{d}s ,\quad
		\text{s.t.} \quad \eqref{eq:lethargycone}-\eqref{eq:Eb0}.
	\end{equation*}
	\normalsize
\end{prob}
As mentioned in the previous subsections, there are two non-convexities in constraints~\eqref{eq:ratioconstraint} and \eqref{eq:motorconstraint}, in the form of a multiplication or division by the vehicle's velocity. To bypass these nonlinearities, we introduce an iterative approach based on the two following problems.

\begin{prob}[Speed-independent Convex Problem]\label{prob:SIM}
	\emph{The minimum-lap-time control strategies for a speed-independent EM model are the solution of the following SOCP:}
	\small
	\begin{equation*}
	 \underset{x,u}{\text{min}}  \int_{0}^{S} {\dtds(s)}\mathrm{d}s , \quad 
	 \text{s.t.} \quad \eqref{eq:lethargycone}-\eqref{eq:Ekinperiod},
	(\ref{eq:ratiolimit})
	-\eqref{eq:EMcone},
	(\ref{eq:torquelimit})-(\ref{eq:Eb0}).\\
	\end{equation*}
	\normalsize
\end{prob}

\begin{prob}[Speed-dependent Convex Problem]\label{prob:SDM}
	\emph{Given a velocity profile, the optimal state trajectories and control strategies are the solution of the following SOCP:}
	\small
	\begin{equation*}
	\begin{aligned}
	& \underset{x,u}{\text{min}} & & \int_{0}^{S} {\dtds(s)}\mathrm{d}s, \quad
	\text{s.t.} \quad \eqref{eq:lethargycone}-\eqref{eq:Ekinperiod}, 
	(\ref{eq:ratiolimit}),
	(\ref{transmissionforcebalance}),
	(\ref{eq:torquelimit})-(\ref{eq:Eb0})\\
	& & & F_{\mathrm{dc}}(s) \geq F_{\mathrm{m}}(s) +  \left( {{x}^\top(s) Q x(s)}\right)/ {\bar{v}(s)}\\
	& & & {\omega_{\mathrm{m}}(s)} = {\gamma(s)} \cdot {{\bar{v}(s)} \cdot \gamma_{\mathrm{fd}}}/{r_{\mathrm{w}}}.
	\end{aligned}
	\end{equation*}
	\normalsize
\end{prob}

As an initial assumption, we solve Problem~\ref{prob:SIM},
and then use the resulting velocity profile as a parameter value to solve Problem~\ref{prob:SDM}.
As shown in Algorithm~\ref{alg:iter}, we iteratively solve Problem \ref{prob:SDM} until the velocity profile of the current solution coincides with the velocity of the previous iteration up to a tolerance $\varepsilon_v>0$. 

\begin{algorithm}[t!]
	\caption{Iterative solving procedure}
	\label{alg:iter}
	\begin{algorithmic}
		\State $\bar{v}(s)$ $\leftarrow$ Solve Problem 2
		\While{norm($v - \bar{v}$) $\geq \varepsilon_v$  }
		\State  $\bar{v}(s) = v(s)$
		\State  $v(s)$ $\leftarrow$ Solve Problem 3
		\EndWhile
	\end{algorithmic}
\end{algorithm}

\subsection{Discussion}
A few comments are in order.
First, we consider static transmission models with a constant efficiency value. This is in line with current high-level modeling approaches. More detailed loss models will be investigated in future research.
Second, we assume that the cooling system is able to cope with the EM working at full power for a longer period of time. This can be interpreted as a qualifying lap where the full capabilities of the EM can be exploited.
Finally, we solve Problem~\ref{prob:main} with an iterative algorithm based on convex optimization, that can be interpreted as a partial sequential second-order conic programming approach. Unfortunately, the nonlinear nature of the problem hinders us from theoretically guaranteeing global optimality. Yet, the low-sensitivity of the optimal speed profile w.r.t.\ small perturbations (see Fig.~\ref{fig:resultscomparison}) and the coherence of the sensitivity analysis results (see Fig.~\ref{fig:DeltaTgrid}) are promising.

%% file: chapters/results.tex
\section{Results}\label{sec:results}

This section presents numerical results showcasing the optimization framework presented in Section~\ref{sec:methodology}. First, we discuss the numerical solutions of the optimization problem, comparing the behavior of the race cars equipped with a single-gear and a CVT.
Finally, we present a case study investigating the impact of the CVT efficiency and the battery energy available on the achievable lap time.

We discretize the continuous model presented in the previous section with the Euler Forward method with step-length $\mathrm{\Delta} s = \unit[10]{m}$.
We parse the problem with YALMIP~\cite{Loefberg2004} and solve it with the second-order conic solver ECOS~\cite{DomahidiChuEtAl2013}.
Solving one optimization problem typically requires three iterations of Algorithm~\ref{alg:iter}, indicating a high robustness of the proposed approach.
This way, the total computation time is about one minute when using an Intel Core i7-3630QM \unit[2.4]{GHz} processor with \unit[8]{GB} of RAM.

\begin{table}[!t]
	\caption{Vehicle Parameters}
	\label{tab:param}
	\begin{tabular}{l c c c}
		\textbf{Parameter} & \textbf{Symbol} & \textbf{Single-gear} & \textbf{CVT}   \\ \hline
		Efficiency - EM to wheels & $\eta_{\mathrm{fd}}\cdot \eta_{\mathrm{gb}}$  [-]& 0.98 & 0.96   \\
		Total vehicle mass & $m_{\mathrm{tot}}$ [kg] & 1341 & 1395   \\
	\end{tabular}
\end{table}

\subsection{Numerical Results}
\begin{figure}[!t]
	\centering
	\includegraphics[width=\columnwidth]{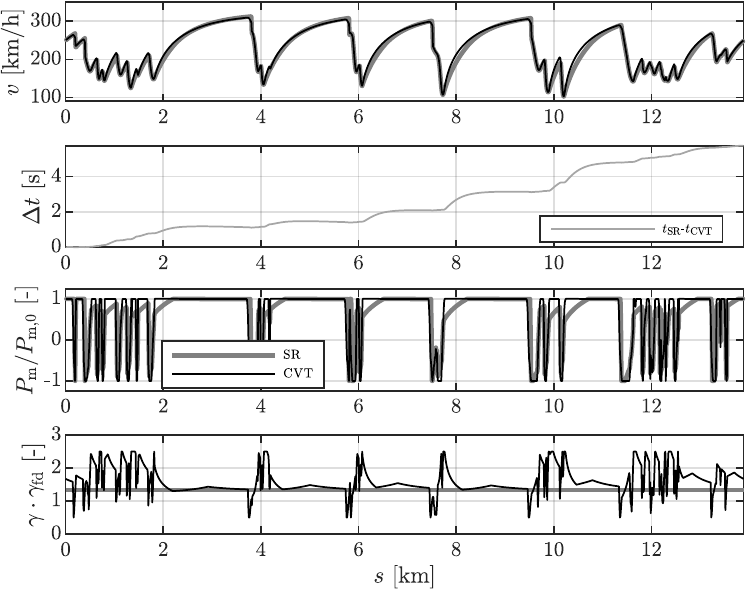}
	\caption{Velocity, accumulated time difference, motor power, and ratio trajectory of both the single-reduction (SR) and the CVT vehicle.}
	\label{fig:resultscomparison}
\end{figure}

In this section, we compute the optimal input and state trajectories for one race lap on the Le Mans racetrack.
The parameters used for the vehicles under consideration are given in Table~\ref{tab:param}. 
Fig.~\ref{fig:resultscomparison} shows the optimal solution for both vehicles for $\mathrm{\Delta} E_\mathrm{b,max} = \unit[75]{MJ}$, with the CVT-equipped one completing the lap in \unit[226.4]{s}: \unit[5.8]{s} faster than the one equipped with an SR.
The SR vehicle is able to reach a higher maximum velocity due to its lower mass and power losses.
Yet, after the apex of relatively low-speed corners, the CVT car can lower the transmission ratio to instantly deploy the maximum EM power while keeping the EM speed constant. Hence it can accelerate faster and gain lap time over the SR vehicle. Moreover, the variable transmission ratio allows a high-efficiency operation of the EM.
Overall, combining these features results in a faster lap time for the CVT car.

\subsection{Parameters Study}
\begin{figure}[!t]
	\centering
	\includegraphics[width=\columnwidth]{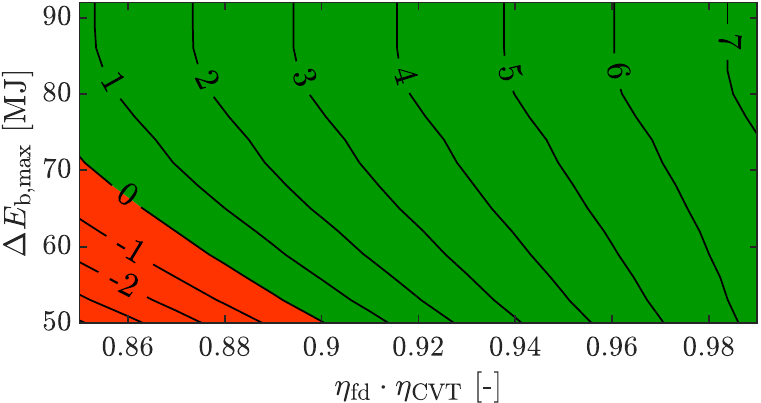}
	\caption{Achievable lap time difference between CVT and SR defined as $\mathrm{\Delta}T = T_{\mathrm{SR}} - T_{\mathrm{CVT}}$ for various EM-to-wheel CVT powertrain efficiency and available battery energy.}
	\label{fig:DeltaTgrid}
\end{figure}

The computational efficiency of the proposed method allows to perform extensive studies and gain insights into the impact of different vehicle parameters on the achievable lap time.
As an example, we vary the CVT-efficiency and the battery energy provided between \unit[85]{\%} and \unit[99]{\%}, and \unit[51]{MJ} and \unit[92]{MJ}, respectively.
This results in a grid of 225 points for which we compute the minimum lap time. The full computation lasted less than four hours, which is in line with the computational time measured for one single problem.
Fig.~\ref{fig:DeltaTgrid} shows the considered CVT car outperforming the SR car in most of the scenarios, with the benefits increasing for larger energy budgets and a higher transmission efficiency.

%% file: chapters/conclusion.tex
\section{Conclusion}\label{sec:conclusion}
In this paper, we devised an optimization framework to compute the minimum-lap-time control strategies for battery electric race cars. Thereby, we first derived a model of the powertrain, including the battery pack, the electric machine, and two types of transmissions, namely a single speed ratio (SR) and a continuously variable transmission (CVT). 
Thereafter, we presented an iterative algorithm that can efficiently compute the minimum-lap-time control strategies using convex optimization in one minute. As a case study, we leveraged the proposed methodology to solve the minimum-time control problem for a battery electric race car on the Le Mans track, whereby we compared the performance of the two transmission technologies and characterized the impact of the energy allowance and CVT efficiency on the achievable lap time. Under a constant transmission efficiency assumption, our study indicated that the lap time of a CVT-equipped electric race car can significantly outperform the one achievable with a SR. These promising preliminary results prompt a more detailed analysis based on high-fidelity models.